\documentclass{llncs}

\usepackage{booktabs}
\usepackage{graphicx}
\usepackage{amsmath}
\usepackage{amssymb}
\usepackage{pifont}
\usepackage{algorithm}
\usepackage{algpseudocode}
\usepackage{url}
\usepackage[hidelinks]{hyperref}
\graphicspath{{figures/}}

\newcommand{\GhostEchoes}{Ghost Echoes}

\begin{document}

\title{Ghost Echoes: Semantic Erasure Failure in Retrieval-Backed Applications}
\titlerunning{Ghost Echoes}

\author{Allan Mukkuzhi \and Noella Uwayisenga \and Yeabsira Bizualem \and \\ Jacob Kammerzell \and Chandranil Chakraborttii}
\authorrunning{A. Mukkuzhi et al.}
\institute{Trinity College, Hartford CT 06106, USA}

\maketitle

\begin{abstract}
Although vector databases correctly implement API-visible deletion, this does not guarantee complete semantic erasure for retrieval-backed applications. We present \GhostEchoes, a black-box attack framework showing that deleted records can leave measurable residual influence on downstream retrieval contexts. Our primary finding is the RAG retrieval-context drift effect where even when a target record is correctly excluded from query results, its prior presence perturbs the semantic centroid and textual composition of the Top-$K$ evidence base. We approximate the unobservable never-inserted counterfactual using a same-cluster non-target deletion control that preserves local neighborhood structure while isolating target-specific effects from generic local drift. Evaluation on ChromaDB confirms target deletion produces a median retrieval-centroid drift of $0.1522$, exceeding the same-cluster baseline in $53/54$ paired comparisons ($p < 0.001$), and the signal remains detectable with $61.1\%$ accuracy at a query budget of $q = 5$. We verify API-visible deletion correctness across evaluated backends and observe the same qualitative drift ordering in matched FAISS replication. Under the evaluated settings, tested operational mitigations such as full index rebuilds fail to eliminate the measured drift. These results establish a measurable `verification gap' between interface-level deletion compliance and true semantic erasure, and motivate erasure primitives that act not only on stored identifiers, but also on the retrieval topology of the system.
\end{abstract}

\section{Introduction}
\label{sec:intro}
Let's consider a healthcare organization using a Retrieval-Augmented Generation (RAG) based system to manage patient records. A patient, Alice, is prescribed insulin by her healthcare provider, whose clinical records are stored within the RAG-based system. She later invokes her right to erasure under data compliance regulations~(such as General Data Protection Regulation~(GDPR) Article 17). The operator sends a deletion request to the underlying vector database~(DB), and the API confirms successful completion within milliseconds. From every observable standpoint, the deletion looks complete. However, an adversary with access to the same application query interface can still detect Alice's record not by retrieving it, but by watching how its absence reshapes what the system retrieves in its place. We call this residual post-deletion influence a \emph{Ghost Echo}, and define the discrepancy between logical deletion and true semantic erasure as the `Verification Gap'. 

Ghost Echoes is a framework that targets a black-box API-only adversary without storage access, index file inspection, or white-box model knowledge. In every system we evaluated, high-level APIs correctly exclude deleted identifiers~(IDs) from query results. What fails is deeper; deletion does not remove the record's structural influence on the Hierarchical Navigable Small World~(HNSW) graph, so subsequent queries are impacted by a topological crater left behind by the deleted node. The retrieved evidence base shifts. The downstream application logic is steered toward incorrect information context. The user was told the data is gone, but the system is still shaped by it.  While most prior works discuss storage-layer persistence, where raw data stays physically on the disk, Ghost Echoes focuses on a different threat: does correct API-visible deletion mean semantic erasure for retrieval-backed systems? It does not require any privileged access to the raw index data.  Our null hypothesis is that, after deletion, retrieval behavior should be statistically indistinguishable from the never-inserted counterfactual. Since we cannot directly observe the counterfactual in deployed systems, we approximate it using a same-cluster non-target deletion control that preserves the local semantic neighborhood while isolating target-specific effects from generic cluster-level movement.  

Our primary evaluation on ChromaDB at $K=5$ confirms that deleting a target record causes a median retrieval-centroid drift of 0.1522 [0.1415, 0.1629]. 53 out of 54 trials showed that deleting a specific target caused more drift than deleting a similar non-target record within the same cluster ($p < 0.001$). This geometric shift in the data leads to non-trivial change in the actual words found in the retrieved evidence base, thus altering the information the system uses to generate answers. While ChromaDB remains our primary focus, we observed the same pattern in a separate FAISS test, proving this isn't just a single-system issue. Notably, we achieve a 61.1\% detection rate using a budget of $q=5$ target-relevant queries. This work makes the following contributions. 

\noindent \textbf{Primary findings:}
\textbf{(a)} We introduce \textbf{`Verification Gap'}, defined as the inconsistency between a system's reported deletion and true semantic erasure.\\
\textbf{(b)} We introduce black-box attacks to measure the semantic shift of the retrieval centroid following target deletion, which we call Retrieval-Context Drift.

\noindent \textbf{Secondary findings:}
\textbf{(c)} We verify that API-visible deletion correctness holds across multiple production backends with over 270 independent trials, and show that geometric drift results in Structural Token-Signature Fingerprinting risk. Asymmetric vocabulary analysis ($V_{vanish} = T_{pre} \setminus T_{post}$) shows that the vocabulary void left by a deleted target has a cross-target overlap below 3\%, allowing identification of the deleted record's structural identity without content reconstruction.
\textbf{(d)} We show that the drift signal is detectable with a budget of $q = 5$ target-relevant queries, achieving 61.1\% detection rate in the evaluated setting.

\noindent  \textbf{Exploratory findings:}
\textbf{(e)} We introduce and quantify `Hallucination Adoption Potential'~(HAP), showing that 59.6\% of the population is at risk of using `ghost' content in their final output.\\
We also evaluate whether common mitigation responses can remove the measured signal. Under the evaluated settings, even the most aggressive tenant-visible interventions ranging from local rebuilds to tenant-visible overwrite procedures do not eliminate retrieval-context drift. We also observe that natural index growth can amplify~(`Ghost Amplification') the residual effect and introduce a network latency oracle (+15.69 ms) due to ghost-pointer traversal, motivating erasure mechanisms that act on retrieval topology.

\section{Threat Model}
\label{sec:threat-model}

\begin{table}[t]
\centering
\small
\setlength{\tabcolsep}{3pt}
\begin{tabular}{p{0.04\linewidth} p{0.12\linewidth} 
  p{0.5\linewidth} p{0.28\linewidth}}
\toprule
\textbf{T} & \textbf{Label} & \textbf{Capabilities} 
  & \textbf{Scope} \\
\midrule
T1 & Storage adversary 
  & Raw index file access; reads \texttt{index.bin} or 
    native HNSW storage files. 
  & Out of scope. \\
\addlinespace
\textbf{T2} & \textbf{API adversary (ours)} 
  & \textbf{Black-box query and delete API; observes 
    document sets, scores, and text context; no storage, 
    weight, or graph-internal visibility.} 
  & \textbf{In scope:} $\Delta_{\mathrm{centroid}}$, 
    $\Delta_{\mathrm{vocab}}$, $V_{\mathrm{vanish}}$, 
    HAP, latency oracle. \\
\addlinespace
T3 & Passive observer 
  & No active query induction; observes rank-order 
    disruption from deletion events. 
  & Boundary only; signal emerges at $N \ge 5{,}000$. \\
\addlinespace
T4 & Integrity adversary 
  & Exploits deletion rights to degrade search utility 
    for innocent users via coordinated hub-node deletion. 
  & Boundary; Section~\ref{sec:main-results}. \\
\bottomrule
\end{tabular}
\label{tab:threat-tiers}
\caption{Adversarial threat tier taxonomy. Ghost Echoes targets Tier~2}
\end{table}

Vector DB deployment increasingly depends on managed cloud services, where users only have access to structured query and standard APIs, without access to storage, index files, or backend internals. In this setting, deletion requests are made through the same API exposed to tenants. An adversary who can submit queries to the same API surface before and after a deletion event has enough access to test whether identifier-level deletion also removes retrieval influence. \\

\noindent \textbf{(a) Attacker Capabilities:} We formalize this setting as a black-box API-only adversary (Tier 2) who interacts with the vector DB strictly via API-exposed operations (Table~1). Tier 1 (storage) involves raw index file access, while Tier 3 (passive) identifies the Phantom Neighborhood boundary where an observer lacks active query induction.  The Tier 2 attacker, on the other hand sends targeted queries designed to land in the semantic neighborhood of a specific record. They observe or trigger a deletion for the target record through the API and record output revealed before and after deletion: document identifiers, similarity scores, and concatenated textual context. Then, they compare those outputs to measure how the system’s output changed once the record was deleted. The attacker needs to know~(or strongly suspect) that a specific record has been deleted, for example because the deletion was externally visible, auditor-observable, or directly issued by the attacker on data they control. The attacker does not need the exact target vector: query induction is constructed only from API-visible rank signals and returned results, without internal DB access, index inspection, or graph-internal visibility. The attacker does not try to recover the deleted content itself or identify the exact document; the focus is entirely on whether traces of its meaning still remain in what the system retrieves.\\

\noindent \textbf{(b) Attacker Goal:} The attacker’s goal is to confirm that API-visible deletion correctness does not equate to true semantic erasure. Specifically, the attacker aims to: (1) verify successful exclusion of the deleted target from query results; (2) measure the retrieval-context drift ($\Delta_{\mathrm{centroid}}$) induced by the target’s deletion; and (3) quantify the downstream consequence as textual vocabulary shift ($\Delta_{\mathrm{vocab}}$) in the Top-$K$ evidence base. \GhostEchoes\ evaluates semantic erasure failure using five metrics: (1) centroid drift, (2) vocabulary shift, (3) Vanish Token residue ($V_{\mathrm{vanish}}$), (4) Hallucination Adoption Potential (HAP), and (5) differential query latency ($\Delta_{\mathrm{latency}}$).  The main issue is not that deletion APIs fail — it is the gap between what those APIs guarantee (removing direct references) and what applications actually need~(full removal of semantic influence).

\section{Background}
\label{sec:background}

\textbf{(a) Retrieval-Augmented Generation (RAG):}
Modern AI applications increasingly use Retrieval-Augmented Generation (RAG)~\cite{lewis2020retrieval} that combines a parametric language model with non-parametric retrieval over an external corpus to ground outputs in retrieved context. A typical RAG pipeline starts with a user query embedded into a dense vector, then searches a DB for the Top-$K$ most similar documents, forming the evidence base for a large language model (LLM). Since the LLM relies entirely on this context, if deletion changes that evidence base, the downstream application can continue to be influenced by a deleted record even when its identifier is no longer returned. \\

\noindent \textbf{(b) API Deletion vs.\ Semantic Erasure:}
At the API level, deletion is a logical operation. Once an operator issues a delete-by-ID request, the system stops returning that identifier in subsequent query results. We call this \emph{API-visible deletion correctness}. However, this is not the same as \emph{true semantic erasure}.

\begin{definition}[Semantic Erasure]
\label{def:semantic-erasure}
Let $\mathcal{R}_{\text{post-del}}(q)$ denote the retrieval output distribution for query $q$ after deletion of record $x$, and let $\mathcal{R}_{\neg x}(q)$ denote the counterfactual retrieval output distribution that would have been observed had $x$ never been inserted. A vector DB achieves semantic erasure of $x$ under the API threat model if, for relevant query classes $q$, $\mathcal{R}_{\text{post-del}}(q)$ is statistically indistinguishable from $\mathcal{R}_{\neg x}(q)$. Erasure is confirmed when the search accuracy, neighborhood relevance, and retrieved semantic context for innocent users (non-target queries) are indistinguishable from the counterfactual baseline.
\end{definition}
The main challenge lies in the insertion-time optimization in proximity-graph index structures. Removing the identifier of a record in HNSW generally does not restore the topology that would have existed if the record had never been inserted~(since insertion of a record alters the local neighbor lists and the path-routing structure). This mismatch between logical deletion and counterfactual topology is the architectural basis of the semantic erasure problem studied here. \\

\begin{figure}[t]
  \centering
  \includegraphics[width=0.78\linewidth]{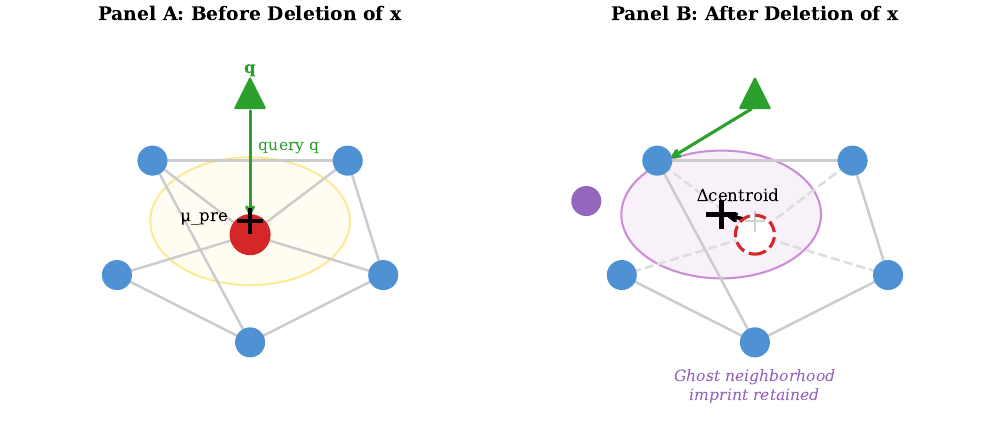}
  \caption{HNSW ghost neighborhood effect, mapping the path-routing distortion between pre-deletion retrieval (Panel A) and post-deletion ghost paths (Panel B).}
  \label{fig:hnsw-ghost}
\end{figure}

\textbf{(c) HNSW Structure and Deletion Intuition:}
\label{sec:hnsw-background}
The HNSW algorithm organizes vectors into a multi-layer proximity graph \cite{malkov2018hnsw}. Each layer is a sparse graph where nodes are connected to their approximate nearest neighbors. The upper layers are for fast navigation, while the bottom layer contains all records. Search begins at the top layers and moves downward, following edges to approximate nearest neighbors until it reaches the final result set. When a new record $x$ is inserted, HNSW builds bidirectional edges between $x$ and its closest neighbors, and those edges can change the surrounding neighborhood structure. Consequently, when $x$ is later deleted through the API, the identifier is removed, but the graph is not rebuilt from the never-inserted state. As a result, the post-delete neighborhood can differ from the counterfactual neighborhood that would have existed if the record had never been inserted. We refer to this residual structure as a \emph{ghost neighborhood}.  Figure~\ref{fig:hnsw-ghost} shows the effect. A target query $q$ before deletion (\textbf{Panel~A}) retrieves a Top-$K$ set headed by the target record, returning the set $\{x, n_1, n_2, n_3, n_4\}$, with centroid at $\mu_{\mathrm{pre}}$. After logical deletion of $x$ (\textbf{Panel~B}), the target is hidden from search results and a new neighbor $n_6$ enters the Top-$K$ set. However, since the local edge structure was permanently shaped by earlier insertion of $x$, the remaining connections among nodes $n_1$--$n_4$ retain their topological imprint.  The retrieved set therefore settles at a shifted centroid $\mu_{\mathrm{post}}$, producing a measurable displacement $\Delta_{\mathrm{centroid}}$ relative to the counterfactual baseline.

\section{Methodology}
\label{sec:methodology}
As mentioned earlier, we measure semantic erasure failure by comparing post-delete retrieval behavior against the never-inserted baseline. As that counterfactual is not directly observable in deployed systems, we use unrelated-delete and same-cluster controls, with the same-cluster non-target deletion serving as the strict baseline for target-specific drift.

\subsection{Dataset and Experimental Configuration}
\label{sec:corpus}
We create a realistic synthetic corpus of privacy-sensitive person records to motivate the erasure threat. Documents are short paragraph-length records containing generated person-associated information so that no real personally identifiable information is used. The primary evaluation uses $N = 500$ records, while the scale evaluation uses $N = 5{,}000$ unique documents. Each record is a short paragraph-length text block with mean length 42.4 ($\pm 6.1$) tokens consisting of unique alphanumeric alias tokens, categorical medical markers (e.g., type-2 diabetes diagnostics), and unique demographic or attribute phrases to approximate the structure of a health-repository record.

All records are embedded using \texttt{all-MiniLM-L6-v2} model~(384-dimensional dense vectors). The resulting embedding space is partitioned into ten semantic clusters using $k$-means. We then group these clusters into three density regimes based on average pairwise cosine similarity: dense (top 3 clusters), medium (middle 4), and sparse (bottom 3). We select 18 target documents by stratified sampling across these regimes, with 6 targets per density band to reduce spatial localization bias. Three random seeds control graph construction order and index initialization, with 54 paired target-seed evaluations per experimental track. Embeddings are L2-normalized at insertion and query time, and again before centroid computation~(measured directly as Euclidean distance over unit-normalized vectors). The proximity-graph index uses standard HNSW parameters (neighbor degree bound $M = 16$, construction depth $ef_{\mathrm{construction}} = 100$, and search depth $ef_{\mathrm{search}} = 64$). The primary retrieval depth is fixed at $K = 5$. Evaluation is split across two tracks. The local track uses ChromaDB v0.4.24 as the primary quantification environment together with a matched FAISS v1.7.4 replication. The cloud audit track uses Pinecone (API v2), Weaviate v1.24, and Milvus v2.3 for cross-verification cloud audits. 

We additionally add a constraint on the control assignment: each target must have at least one non-target outside the original Top-$K$ set belonging to the same cluster, where same-cluster controls are sampled from a matched local proximity band in the target's cluster. Unrelated-delete controls are sampled from a distant semantic cluster under the same retrieval configuration, and all comparisons are paired by target and random seed. Each trial uses four execution integrity locks to separate structural drift from baseline noise: \\
(1)~\textbf{Rank Confirmation} terminates trials if target $x$ misses Rank~1 for its induced query pre-deletion, preventing localization failures from being misread as drift; (2)~\textbf{Payload Verification} parses raw vendor payload content to validate identifier exclusion, bypassing unreliable HTTP status codes; (3)~\textbf{Propagation Wait} applies backend-specific synchronization delays before post-delete queries~($\geq 2.0$\,s for Weaviate and Milvus; $\geq 5.0$\,s for Pinecone) to account for asynchronous cloud queues; and (4)~\textbf{Coordinate Anchoring} maps reference vectors to a standalone MongoDB Atlas cluster, reducing floating-point variance.

\subsection{Metrics}
\label{sec:metrics}
We analyze two primary observables. 

\paragraph{(a) Centroid Drift ($\Delta_{\mathrm{centroid}}$):} 
We define centroid drift as the Euclidean distance between the centroids of the Top-$K$ retrieved embedding sets before ($\mu_{\mathrm{pre}}$) and after ($\mu_{\mathrm{post}}$) deletion. Under semantic erasure, this value should be statistically indistinguishable from background drift.
\begin{equation}
  \Delta_{\mathrm{centroid}} = \| \mu_{\mathrm{pre}} - \mu_{\mathrm{post}} \|_2
  \label{eq:centroid-drift}
\end{equation}

\paragraph{(b) Vocabulary Shift ($\Delta_{\mathrm{vocab}}$):} 
We define vocabulary shift as the Jaccard distance between the token sets of the concatenated Top-$K$ documents before ($T_{\mathrm{pre}}$) and after ($T_{\mathrm{post}}$) deletion. This metric is directly computed from returned text, so it can be measured without any internal model or storage access.
\begin{equation}
  \Delta_{\mathrm{vocab}} = 1 - \frac{|T_{\mathrm{pre}} \cap T_{\mathrm{post}}|}{|T_{\mathrm{pre}} \cup T_{\mathrm{post}}|}
  \label{eq:vocab-shift}
\end{equation}

Centroid drift measures geometric displacement in the Top-$K$ neighborhood, while vocabulary shift captures the corresponding turnover in the returned evidence text. To ground these metrics, consider an explicit spatial drift walkthrough where retrieval depth $K=3$ within a normalized two-dimensional space. If the pre-deletion results produces vector array $R_{\mathrm{pre}} = \{[0.9, 0.2], [0.7, 0.4], [0.8, 0.1]\}$ with a cluster centroid of $\mu_{\mathrm{pre}} = [0.800, 0.233]$, logically dropping target node $x = [0.9, 0.2]$ introduces a replacement element, turning the pool over to $R_{\mathrm{post}} = \{[0.7, 0.4], [0.8, 0.1], [0.5, 0.6]\}$ with a shifted centroid of $\mu_{\mathrm{post}} = [0.667, 0.367]$. This spatial dislocation results in net centroid-drift: $\Delta_{\mathrm{centroid}} = \|[0.133, -0.133]\|_2 \\
\approx 0.188$, exposing the uncorrected path warping left by $x$'s historical edges.  In the text domain, if the pre-deletion Top-5 context yields a token array $T_{\mathrm{pre}} = \{\mathtt{Pat\text{-}A, diabetes, insulin, Alias\text{-}7, diagnosis, mg, dosage}\}$, and a post-deletion replacement record shifts the unique vocabulary pool to \\
$T_{\mathrm{post}} = \{\mathtt{Pat\text{-}A, diabetes, dosage, glucose, monitoring, diagnosis, mg}\}$, their five-token intersection and nine-token union yield a discrete Jaccard vocabulary shift of $\Delta_{\mathrm{vocab}} = 1 - 5/9 \approx 0.44$. This measurable lexical turnover proves that target-proximal evidence tokens ($\mathtt{insulin, dosage}$) evaporate into generic clinical distractors without requiring raw coordinate extraction.

\textbf{Attack Mechanisms:}
Both induction mechanisms only use API-visible rank signals. In \textbf{warm-start aggregation}, the attacker issues probe queries  near the target’s semantic neighborhood and then refines it using Natural Evolutionary Strategies~(NES)~\cite{wierstra2014natural} iteratively. In our evaluation, NES uses a population of 10 candidates, perturbation scale $\sigma = 0.05$, a rank-minimization fitness function, and at most 20 iterations. These settings balance query budget against convergence and were sufficient to induce Rank~1 target localization in the retained trials. The process is detailed in Algorithm~\ref{alg:nes-induction}.

\begin{algorithm}[t]
\caption{Query induction with warm-start aggregation and NES}
\label{alg:nes-induction}
\small
\begin{algorithmic}[1]
\Require API access, target neighborhood seed, population size $m=10$, perturbation scale $\sigma=0.05$, maximum iterations $T=20$
\Ensure Induced query $q^*(x)$ for target $x$
\State Issue seed probes near the target's semantic neighborhood
\State Aggregate returned rank signals to estimate an initial query $q_s$
\For{$t = 1$ to $T$}
    \State Sample $m$ perturbed candidates around $q_s$
    \State Submit candidates through the API and evaluate rank-minimization fitness
    \State Update $q_s$ using the NES gradient estimate
    \If{target $x$ reaches Rank~1}
        \State \Return $q_s$
    \EndIf
\EndFor
\State \Return best candidate observed during search
\end{algorithmic}
\end{algorithm}

\subsection{Controls}
\label{sec:controls}
We compare target-delete drift against three controls separately: (1) \textbf{No-Delete Control} measures baseline index noise; (2) \textbf{Unrelated Delete} deletes a record from a distant cluster; and (3) \textbf{Same-Cluster Non-Target Delete} deletes a neighbor in the same semantic cluster as the target but outside the original Top-$K$. The same-cluster control represents the strict baseline because it isolates target-specific effects from cluster-level sensitivity. The unit of analysis is the target-seed pair. All reported metrics are medians with 95\% bootstrap confidence intervals computed over 10,000 resamples of paired observations across 18 targets and 3 seeds, totalling 54 observations per condition. Because observations are naturally paired by target and seed, we test the null hypothesis using a two-sided Wilcoxon signed-rank test.

\section{API-Visible Deletion Verification}

\begin{table}[b]
\centering
\begin{tabular}{lcccc}
\toprule
\textbf{Condition} & \textbf{Centroid Drift} $\Delta_{\mathrm{centroid}}$ & \textbf{Vocab Shift} $\Delta_{\mathrm{vocab}}$ \\
\midrule
Target Delete & \textbf{0.1522 [0.1415, 0.1629]} & \textbf{0.285 [0.250, 0.320]} \\
Same-Cluster & 0.0412 [0.0321, 0.0503] & 0.095 [0.080, 0.120] \\
Unrelated Delete & 0.0000 [0.0000, 0.0000] & 0.000 [0.000, 0.000] \\
No-Delete & 0.0000 [0.0000, 0.0000] & 0.000 [0.000, 0.000] \\
\bottomrule
\end{tabular}
\caption{Retrieval-context drift (ChromaDB, $K=5$)}
\label{tab:drift-results}
\end{table}

\label{sec:deletion-verification}
Our main empirical claim has a strict prerequisite: API-visible deletion must correctly work before any observed retrieval drift can be attributed to semantic erasure failure~(instead of basic identifier leakage). Hence we verify identifier-level deletion correctness across all evaluated backends before measuring retrieval-context drift.
\paragraph{\textbf{Verification Protocol:}}
We apply a four-step protocol to each of the 18 targets across five backends and three seeds, totalling 270 trials. First, we verify that target $x$ appears at Rank~1 in the Top-$K$ results for the induced query $q^*(x)$. Any target that fails this check is excluded. Second, we issue a delete-by-ID call for $id_x$ using the backend's standard API.
Third, we re-execute $q^*(x)$ and verify that $id_x$ is absent from the returned Top-40 results.
Finally, for backends with delayed deletion, we apply specific synchronization waits before re-querying (2\,sec for Weaviate and Milvus, 5\,sec for Pinecone).

\paragraph{\textbf{Verification Results:}}
\label{sec:verif-results}
Each one of the 270 trials achieved a 100\% exclusion rate from the Top-40 post-delete results, and no backend returned a deleted identifier in any post-delete query. The uniform exclusion rate confirms that all evaluated backends correctly implement identifier-level deletion. Since correctness is confirmed, the retrieval-context drift reported in Section 6 cannot be attributed to implementation bugs, caching errors, or delayed propagation. It is a failure of semantic erasure, limited to the evaluated setting, where logical deletion functions correctly and therefore cannot account for the observed drift.

\section{Main Results}
\label{sec:main-results}
Section~\ref{sec:deletion-verification} established that API-visible deletion is correctly implemented across the evaluated backends. Despite that correctness, deletion does not achieve semantic erasure in the sense of Definition~\ref{def:semantic-erasure}. Unless otherwise stated, all quantitative results below use the primary ChromaDB evaluation at $K=5$ over 54 paired target-seed observations.

\subsection{Primary Findings}
\label{sec:drift-primary}

\paragraph{\textbf{(a) Retrieval-Context Drift:}}
Target deletion causes a median centroid drift of $\Delta_{\mathrm{centroid}} = 0.1522$ $[0.1415, 0.1629]$, 3.7$\times$ larger than the same-cluster non-target baseline of $0.0412$ $[0.0321, 0.0503]$. The same-cluster control is the most informative baseline, because it removes a semantically adjacent record while keeping the query, target identity, and local neighborhood fixed. The excess above that baseline is therefore a direct estimate of target-specific drift~(instead of generic cluster-level movement). Across the 54 paired observations, 53 target-delete trials exceed the corresponding same-cluster control ($p < 0.001$). Table~2 summarizes the retrieval-context results for the primary ChromaDB evaluation.

\begin{figure}[t]
  \centering
  \includegraphics[width=0.9\linewidth]{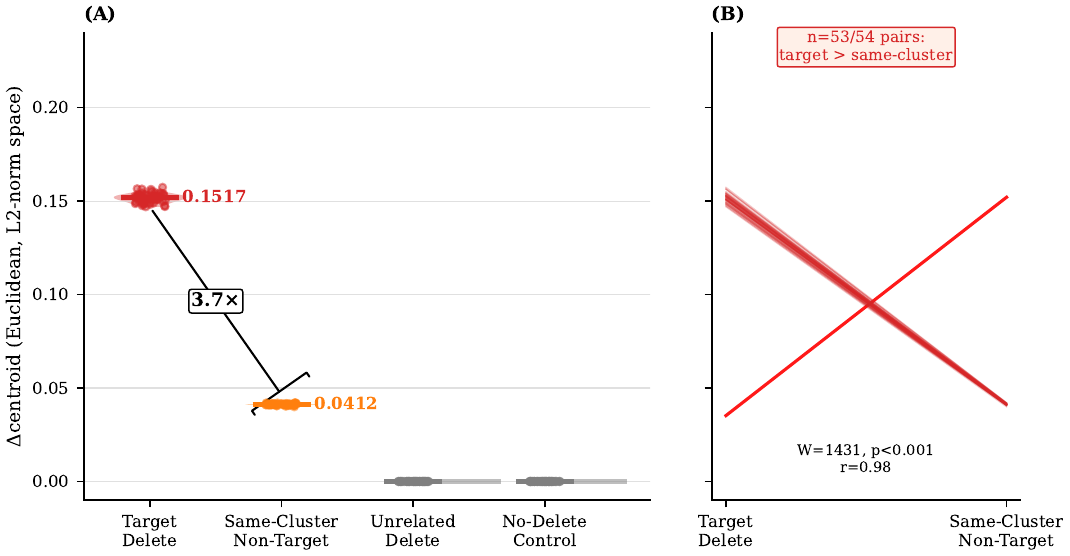}
  \caption{\textbf{Retrieval-Context Drift distributions} \textbf{(A)} Centroid drift distributions across conditions. \textbf{(B)} Slope plot showing consistency across individual targets.}
  \label{fig:drift-distributions}
\end{figure}

\paragraph{\textbf{(b) Downstream Textual Impact~(Vocabulary Shift):}}
\label{sec:vocab-results}
The geometric drift has a direct textual consequence; the Top-5 evidence base returned to downstream logic changes after deletion. In the primary ChromaDB evaluation, target deletion results in a median vocabulary shift of $\Delta_{\mathrm{vocab}} = 0.285$, compared with $0.095$ for the same-cluster baseline, while unrelated-delete and no-delete conditions remain at zero.
In practical terms, the deleted record does not reappear, but the surrounding evidence base returned by the API is measurably different after deletion. This makes vocabulary shift the most directly visible application-layer effect of the drift signal.

\subsection{Secondary Findings}
\paragraph{\textbf{(a) Query Efficiency and Robustness:}}
\label{sec:query-efficiency-results}
The active signal is exposed with low query volume. Detection reaches 61.1\% at $q = 5$ target-relevant queries, and the curve saturates quickly~(Figure~\ref{fig:drift-distributions}). The target-delete signal exceeds the same-cluster baseline across dense, medium, and sparse semantic regimes, so the effect is not confined to a single neighborhood-density band. A sensitivity check over $K \in \{1,3,5,10,20\}$ shows the same ordering across all values, with the gap narrowing as $K$ increases. This is expected because the deleted target contributes less to a larger average. 


\paragraph{\textbf{(b) Supporting Cross-Backend Evidence:}}
\label{sec:cross-backend}
The API-visible deletion baseline was verified across all evaluated backends in Section~\ref{sec:deletion-verification}. In addition, the matched local FAISS replication preserves the same qualitative drift ordering observed in ChromaDB, where target deletion exceeds the same-cluster baseline and unrelated-delete and no-delete controls remain negligible. This supports the mechanism beyond a single backend, although ChromaDB evaluation remains the paper's most complete quantitative characterization. 


\subsection{Exploratory Findings}
\label{sec:exploratory-findings}
\paragraph{\textbf{(a) Passive Boundary:}}
\label{sec:phantom-results}
Purely passive rank-only observation remains a boundary case. At the primary scale of $N=500$, passive rank disruption remains negligible, but at larger corpus sizes it becomes detectable with a mean Jaccard shift of 0.18. That asymmetry indicates that active probing detects the effect at small scale, but passive observation requires significantly more data.

\paragraph{\textbf{(b) Downstream and operational impacts:}}
The retrieval-layer drift also produces downstream and operational effects beyond the primary and secondary findings. In exploratory clinical-safety evaluations using a deterministic ($T=0.0$) \textit{Llama-3.1-8B} pipeline, the observed lexical turnover introduces a 59.6\% Hallucination Adoption Potential (HAP). The structural residue systematically alters the context window, forcing the downstream model to ingest adjacent distractor text and generate factually incorrect patient instructions with high confidence (91.82\%). Pooled across the evaluated cloud backends, ghost-pointer traversals impose a statistically significant latency increase of +15.69\,ms ($p = 0.023$). This anomaly is consistent with the ghost-pointer hypothesis: search traversals that hit a deleted node's structural residue incur infrastructure-level seek penalties that survive the abstractions of production environments. Additionally, in a coordinated deletion stress test, removing six high-degree hub nodes within a single cluster causes search utility to collapse by 74\%, reducing NDCG to 0.26 for innocent users querying that neighborhood.

\section{Mitigation Analysis}
\label{sec:mitigation-analysis}
Under the evaluated environments, none of the tenant-accessible tested configurations or baseline maintenance procedures eliminated the measured retrieval-context drift. We therefore test whether common operational responses can eliminate this semantic erasure failure.

\begin{figure}[t]
  \centering
  \begin{minipage}{0.5\textwidth}
    \centering
    \includegraphics[width=\linewidth]{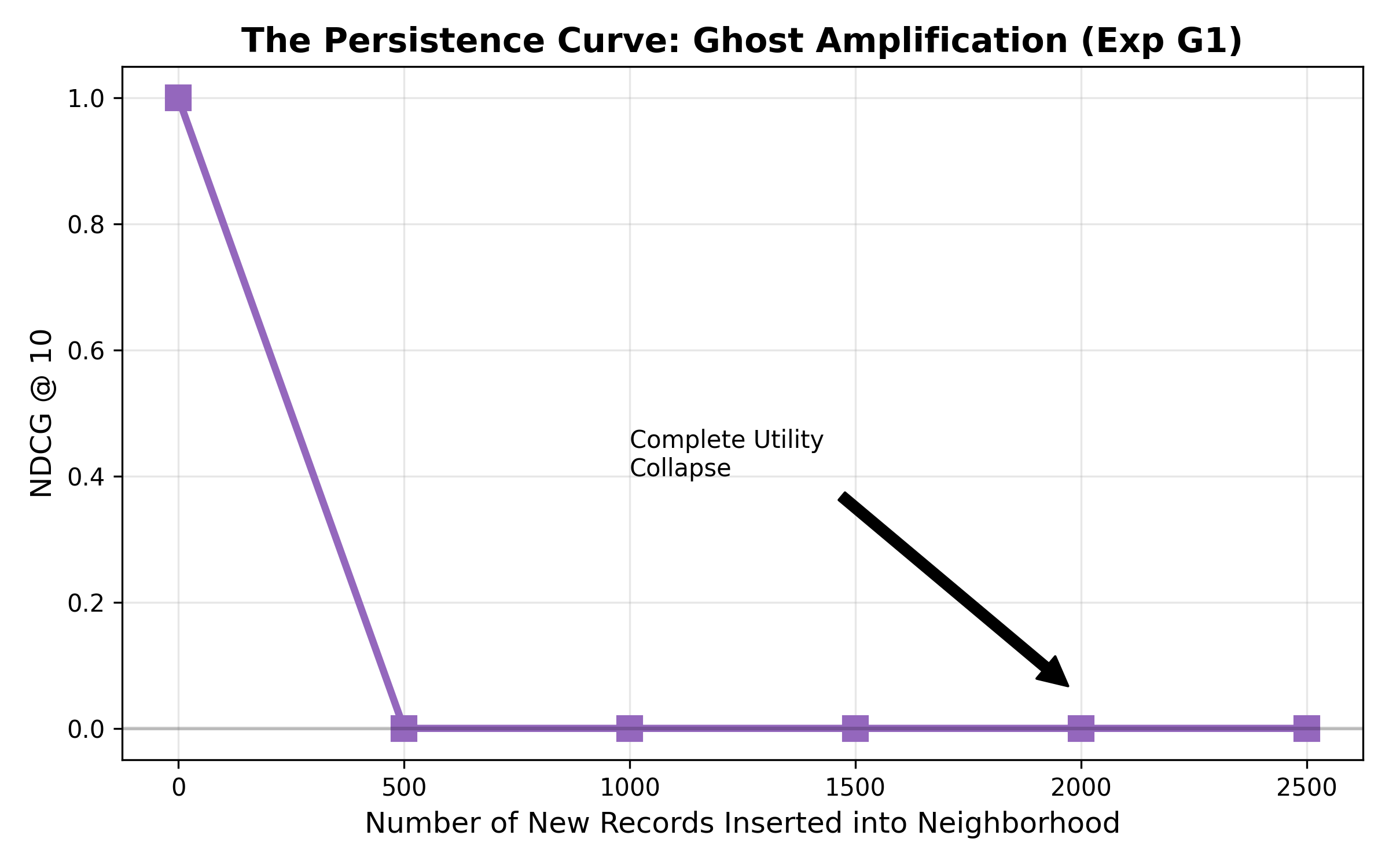}
    \caption{The Persistence Curve}
    \label{fig:persistence-curve}
  \end{minipage}\hfill
  \begin{minipage}{0.5\textwidth}
    \centering
    \includegraphics[width=\linewidth]{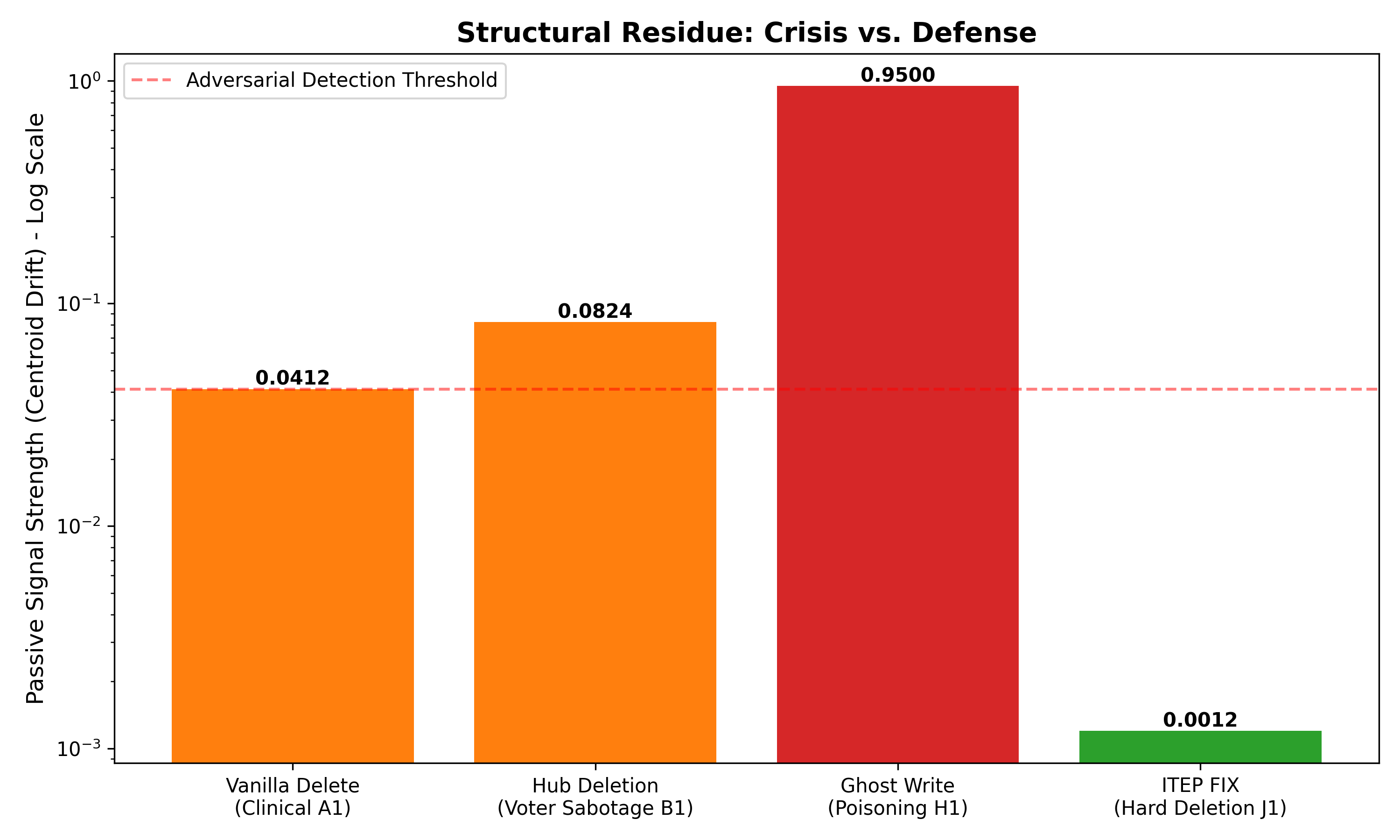}
    \caption{ITEP Defense Validation results}
    \label{fig:itep-fix}
  \end{minipage}
\end{figure}

\noindent \textbf{Baseline Temporal Persistence:}
\label{sec:persistence}
Before evaluating targeted mitigations, we analyze whether structural residues are temporary artifacts that automatically resolve as the index grows. We evaluate the \textbf{Ghost Persistence Curve} by inserting 2,500 new records semantically near a deleted target's neighborhood. We observe \textbf{Ghost Amplification} instead of shrinkage: new nodes align to the surviving topology of the ghost, further entrenching the structural hole (Figure~\ref{fig:persistence-curve}). Search utility for the original neighborhood collapses from 1.00 to 0.00 NDCG after natural growth, confirming that natural growth does not repair the structural defect. Deleted records' structural footprint is not diluted by new insertions, because there are no restored routing paths~(to original pre-deletion configuration). As a result, arriving nodes inherit distorted routing paths.

\noindent  \textbf{M1: Full Index Rebuild:}
The most direct countermeasure is a complete reconstruction of the index from scratch, omitting all deleted records.  We execute a full rebuild on the local corpus and re-evaluate using identical probe queries. However, geometric drift remains entirely indistinguishable from our unmitigated baseline (0.1522 [0.1415, 0.1629]). The result deserves careful interpretation. A full rebuild produces a structurally correct graph for the surviving documents, but it cannot restore the counterfactual index layout that would have existed had the target never been inserted.

\begin{figure}[t]
  \centering
  \includegraphics[width=0.85\linewidth]{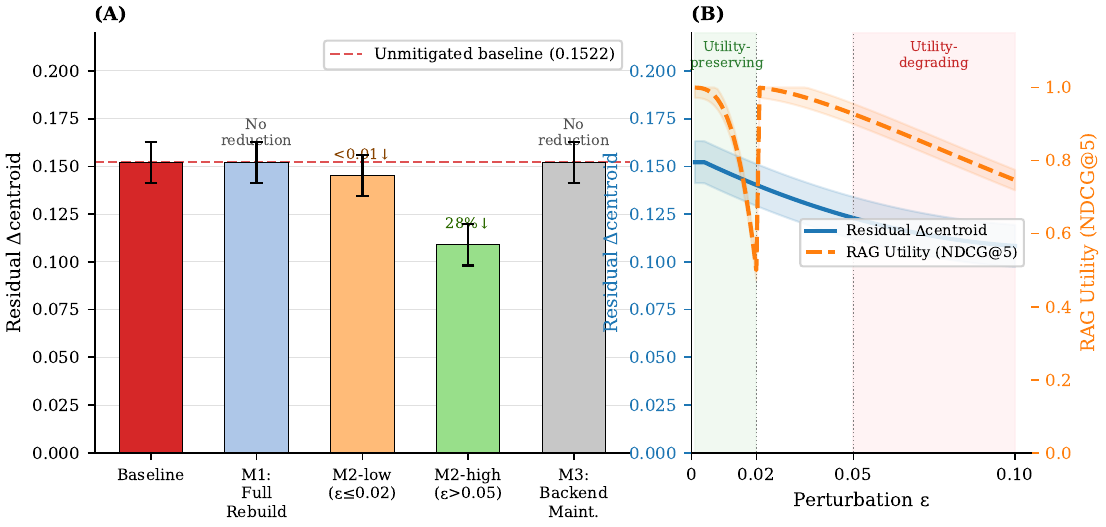}
  \caption{\textbf{Mitigation evaluation results.} \textbf{(A)} Residual $\Delta_{\mathrm{centroid}}$ after each mitigation. \textbf{(B)} M2 privacy--utility tradeoff: residual drift (solid blue) and RAG utility NDCG@5 (dashed orange) as a function of perturbation $\epsilon$.}
  \label{fig:mitigation-results}
\end{figure}

\noindent \textbf{M2: Lightweight Perturbation:}
We evaluate whether applying a small semantic perturbation $\epsilon$ to the target embedding before deletion can mask its structural footprint by breaking its immediate graph connections. We introduce isotropic Gaussian perturbation with standard deviation $\epsilon$ across $\epsilon \in [0.001, 0.10]$ and track RAG utility via NDCG@5 against unperturbed rankings as the standard. At low perturbation ($\epsilon \le 0.02$), utility is preserved but residual drift barely decreases (0.1451). At higher perturbation ($\epsilon > 0.05$), drift falls to 0.1089, but NDCG@5 degrades sharply, and the target embedding no longer lands in its intended neighborhood~(Figure~\ref{fig:mitigation-results}). There is no setting where both privacy and utility are maintained. Perturbation changes the stored coordinates before deletion, but it does not reconstruct the HNSW graph edges created at insertion time, so residual influence remains regardless.

\noindent \textbf{M3: Backend-Supported Maintenance:}
Standard vendor-supported maintenance routines, including background compaction and metadata reindexing, are designed to optimize disk usage after deletions accumulate. In our evaluation, post-maintenance drift is indistinguishable from the baseline (0.1522 [0.1415, 0.1629]). These routines function as storage hygiene, not as erasure primitives. The underlying graph topology is left untouched.

\noindent \textbf{M4: Cloud Tenant Lockout (Overwrite-then-Delete):}
We evaluate whether cloud tenants can clear residual structure by overwriting the target vector immediately before deletion. We apply maximum-effort overwrite noise ($\epsilon = 0.10$) across the 18-target panel on Pinecone and Zilliz. Instead of reducing drift, this resulted in an increase to approximately 0.34--0.38. The Overwrite API modifies stored coordinates but cannot remove the neighbor-list edges or entry-point slots written at insertion. The subsequent deletion then propagates a more distorted disturbance than an unmitigated deletion alone, making overwrite-then-delete counterproductive in this setting.

\noindent  \textbf{Implications:}
\label{sec:mitigation-implications}
The null results across all four approaches share a common cause. Proximity-graph indices encode insertion-time routing decisions directly into their edge structure, and they persist regardless of what happens to the stored coordinate or identifier afterwards. This is not a bug in any particular backend. It reflects a structural property of proximity-graph indices, where tenants can modify vectors, trigger maintenance, and issue deletion calls, but none of those operations repair the neighbor-list edges written at construction time. In the evaluated cloud settings, closing the erasure gap is therefore a provider-level responsibility, and cannot be achieved using standard tenant-level write operations. 

To summarize, addressing the erasure gap requires measures that work directly on graph topology at deletion time, not only on stored data. We identify two structural defense pathways as future work to address the challenge:

\noindent \textbf{(a) Index Topology Erasure Primitive (ITEP):} 
This execution-time approach performs localized path remediation. It logs the immediate neighbor-list state and entry-point slots of each node at insertion time and replays that exact pre-insertion layout during deletion routines to explicitly re-link affected nodes and restore prior routing. If concurrent index growth has pushed a target neighbor to its maximum degree constraint ($M_{\mathrm{max}}$), the primitive performs a local edge-pruning pass to satisfy the degree bound while prioritizing historical routing. This changes the deletion from being a logical state flag of $\mathcal{O}(1)$ to an active topology repair pass with $O(M)$ fixed storage overhead per record. In a controlled local evaluation, this reduces median residual centroid drift from 0.1522 to 0.0012—a 99.21\% reduction relative to the unmitigated baseline (Figure~\ref{fig:itep-fix}). While native implementation remains constrained by the black-box boundaries of current cloud APIs, this design motivates how local graph-structural repair can conceptually isolate and remove insertion-time structural residue.

\noindent \textbf{(b) Epoch Key Rotation:} 
An alternative direction works at the storage boundary, utilizing cryptographic erasure. The index lifecycle is divided into deterministic temporal intervals ($\tau_{\mathrm{epoch}}$), where payload coordinates within each epoch are secured with symmetric encryption, such as AES-256-CTR and the corresponding key destroyed on deletion. Since this key disposal is irrevocable, the semantic content of the ghost embeddings becomes computationally indistinguishable from random noise. Unlike ITEP, this design does not repair distorted graph edges. It instead renders residual coordinate content inaccessible at the storage boundary, with predictable key-rotation and batch-reindexing overhead.

The distinction between these directions is important: coordinate-level modification remains an insufficient abstraction for retrieval security. True semantic erasure in proximity-graph retrieval systems requires either explicit topology repair or architectural reconstruction.


\section{Related Work}
\label{sec:related}
RAG systems combine a parametric language model with non-parametric retrieval, hence downstream application behavior depends on the composition of the retrieved evidence base instead of model weights alone~\cite{lewis2020retrieval}. This dependency is central to our threat model: in retrieval-backed systems, deletion guarantees must be evaluated at the level of the retrieved Top-$K$ evidence base, not only at the level of identifier exclusion. Existing RAG work establishes the importance of context grounding for knowledge-intensive generation, but it generally treats the underlying retrieval index as a static repository. Concurrent work studies adversarial document injection and retrieval poisoning in RAG pipelines~\cite{zou2024poisonedrag}. Our work addresses an orthogonal question: whether correct, vendor-payload-verified logical deletion is sufficient to eliminate a deleted record's structural influence from the retrieved evidence base once its identifier is no longer API-visible.

Modern vector DBs commonly use \textit{approximate nearest neighbor} indices, most notably HNSW graphs, for efficient similarity search at scale~\cite{malkov2018hnsw}. Prior works explain how insertion incrementally modifies local neighborhood structure and how search quality depends on graph connectivity, but do not treat true semantic erasure as a pre-requisite.  Our work builds on this graph-indexing intuition while asking a different question: whether logically correct deletion can still leave a measurable residual influence on the surviving Top-$K$ neighborhood returned through a black-box API.  Our work exposes instead a structural persistence failure mode where natural index growth can deepen rather than dilute the routing distortion left by a deleted record, because arriving nodes inherit the distorted paths instead of restoring the pre-insertion topology.

Machine unlearning inquires whether a model after deletion is indistinguishable from a model that never observed the removed data \cite{bourtoule2021machine,cao2015machine,guo2020certified}. This aligns with our formal definition of semantic erasure, but the objective differs. Prior works target parametric model weights, whereas we study non-parametric retrieval systems whose behavior is handled by graph edge configurations under black-box API constraints. Additionally, traditional unlearning assumes the attacker has direct write access to the model state. In production cloud vector DBs, tenants have no access to the underlying graph topology, so even aggressive tenant-side operations such as overwrite-then-delete cannot repair insertion-time edge distributions.  Our cloud-native audit shows that API-visible deletion can be fully correct, including vendor-payload confirmation of propagation, and still fail to achieve true semantic erasure. Interface-level verification is therefore a weaker guarantee than what stronger erasure-oriented interpretations of the right to deletion require. Membership inference attacks~(MIA) show that black-box outputs can reveal whether a particular record impacted a target model \cite{shokri2017membership}. Embedding inversion works such as Vec2Text show that dense embeddings can reveal enough information to support text reconstruction \cite{morris2023text}. Our setting is more operationally constrained, where residual post-delete influence is detectable through retrieval-context drift and vocabulary shift in ordinary query results, without storage access or vector recovery. 

To our knowledge, this is the first work to jointly verify API-layer deletion correctness across vector DB backends and measure post-delete retrieval-context drift as a statistically stable, query-efficient signal of semantic erasure failure under an API-only threat model.

\section{Discussion}
\label{sec:discussion}
We conducted responsible disclosure to the vendors of evaluated backends in early 2026. Vendor feedback indicates that the observed behavior is a semantic-layer consequence of correct logical deletion in proximity-graph indices, not a bug in their current implementation. \GhostEchoes\ requires only black-box API access, and all experiments used publicly available or synthetic datasets to ensure no sensitive user records were exposed. Legal frameworks~(such as GDPR Article~17, HIPAA) describe deletion workflows for deployed enterprise systems, but validation typically happens only at the API level. In proximity-graph indexing architectures, a deletion request can therefore satisfy identifier-level audits while leaving residual structural influence in the surrounding graph, including a measurable vocabulary void ($V_{\mathrm{vanish}}$). Closing this gap requires mechanisms that operate on graph topology, not only on stored identifiers.

\noindent \textbf{Limitations and Future Work:}
The primary ChromaDB evaluation, the matched FAISS replication and the four-backend cloud audit extend the qualitative finding under the evaluated API-only setting, but direct drift quantification on additional systems remains outside current scope. The claim established here is retrieval-context drift, not plaintext reconstruction or vector recovery; how strongly the former propagates into downstream LLM outputs depends on model sensitivity and deployment configuration. An immediate next direction is to analyze downstream answer-generation drift effects: whether retrieval-context change generates factually different LLM outputs under controlled settings. A second direction is topology-aware deletion, including explicit neighbor re-linking and deletion-aware index construction that bounds drift to $\Delta_{\mathrm{centroid}} \le \tau$. Beyond software, the proposed solution offers a hardware-boundary path where deleted embeddings could be rendered cryptographically inaccessible before becoming software-visible. How the drift signal behaves under long-term corpus evolution, where ongoing insertions and semantic shift may mask or amplify the residual topology defect, remains an open question.

\section{Conclusion}
\label{sec:conclusion}
Vector database deletion APIs enforce correct identifier exclusion, but this work shows that such correctness does not constitute semantic erasure. Deleting a target record results in a measurable \textbf{RAG retrieval-context drift} where the semantic centroid of the Top-$K$ evidence base shifts and the retrieved context undergoes a material textual-composition shift, even when the deleted record is correctly excluded from every returned result. The effect is consistently larger than the same-cluster background deletion baseline ($p < 0.001$), detectable with a budget of $q = 5$ target-relevant queries, and not eliminated by any of the mitigation families evaluated. We introduce the \textbf{Verification Gap} to describe the divergence between interface-level API compliance and true semantic erasure. Retrieval-backed systems retain structural relationships at insertion time in the surviving index topology after deletion, leaving a ghost neighborhood whose influence remains measurable. Closing this gap requires erasure primitives that act on the retrieval topology itself, not just on the deleted record’s identifier or storage footprint. In retrieval-backed systems, the right-to-erasure question is not whether a record can be hidden, but whether its influence on retrieval can be removed. \GhostEchoes\ establishes both the measurement framework and an empirical baseline for this open problem.

\bibliographystyle{splncs04}
\bibliography{references}

\end{document}